\documentclass{article}
\usepackage{spconf,amsmath,amssymb,amsfonts,graphicx,hyperref}
\usepackage{amsthm}
\usepackage{algpseudocodex}
\usepackage{algorithm}
\usepackage{textcomp}
\usepackage{bm}
\usepackage{placeins}
\usepackage{xcolor}
\usepackage{float}
\usepackage{subcaption}
\usepackage[normalem]{ulem}
\usepackage{subcaption}
\usepackage{enumitem}

\def\BibTeX{{\rm B\kern-.05em{\sc i\kern-.025em b}\kern-.08em
   T\kern-.1667em\lower.7ex\hbox{E}\kern-.125emX}}

\usepackage[
 backend=biber,
 style=ieee,
 maxbibnames=99,
 minbibnames=99,
 doi=false,
 isbn=false,
 url=false,
 eprint=false
]{biblatex}

\newcommand{\oldcomment}[1]{}

\newtheorem{proposition}{Proposition}

\DeclareMathOperator*{\Argmin}{\mathrm{Argmin}}

\title{MULTIRESOLUTION BLOCK-COORDINATE PLUG-AND-PLAY ALGORITHM FOR IMAGE RECONSTRUCTION}

\name{\parbox{0.95\textwidth}{\centering
Edgar Desainte-Maréville$^{\star}$ \quad Marion Foare$^{\dagger\star}$ \quad Paulo Gonçalves$^{\ddagger}$ \\
Nelly Pustelnik$^{\mathsection}$ \quad Elisa Riccietti$^{\star}$}}

\address{\parbox{0.95\textwidth}{\small\centering
$^{\star}$ ENS de Lyon, CNRS, Inria, Université Claude Bernard Lyon 1, LIP, UMR 5668, 69342 Lyon cedex 07, France \\
$^{\dagger}$ CPE Lyon, Villeurbanne, 69100, France \\
$^{\ddagger}$ Inria, CNRS, ENS de Lyon, Université Claude Bernard Lyon 1, LIP, UMR 5668, 69342 Lyon cedex 07, France \\
$^{\mathsection}$ CNRS, ENS de Lyon, Laboratoire de Physique, UMR 5672, Lyon, France}}

\begin{document}

\maketitle

\begin{abstract}
Plug-and-Play methods are among the state-of-the-art approaches for image reconstruction, but they do not scale well with the image size. 
We propose a multiresolution block-coordinate Plug-and-Play algorithm that decomposes the image into wavelet blocks and applies a block-wise denoiser network to only a subset of blocks at each iteration, selected through a stochastic Gauss-Southwell-inspired activation rule. 
Numerical results on deblurring show that the proposed method outperforms the standard Plug-and-Play in different degradation regimes, and in particular the adaptive rule dynamically adjusts to the degradation regime to match the optimal activation rule in every context.

\end{abstract}


\section{Introduction}

Physical models for imaging inverse problems have been combined with data-driven denoisers, notably through Plug-and-Play (PnP) algorithms, where the regularization term is implicitly learned through a trained neural network denoiser that replaces a proximal operator~\cite{venkatakrishnan_plug-and-play_2013}.
Numerous PnP variants have since been proposed~\cite{chan_plug-and-play_2017, hurault_gradient_2022}, achieving state-of-the-art performance across a wide range of imaging reconstruction tasks~\cite{kamilov_plug-and-play_2023}.
In this work, we focus on the Forward--Backward PnP (FB-PnP) algorithm, which alternates gradient descent steps on a data-fidelity term, and denoiser steps.
However, FB-PnP scales poorly with the problem dimension
~\cite{kamilov_plug-and-play_2023}.

One classical way of reducing this cost is to use multigrid or multilevel methods, which exploit a hierarchy of scales 
~\cite{nash_multigrid_2000}. Originally designed for smooth optimization, this idea was later applied to non-smooth optimization~\cite{parpas_multilevel_2017} and to image restoration~\cite{lauga_iml_2024, aleotti2026multilevel, lauga2024multilevel}.
More recently, Laurent et al.~\cite{laurent2025multilevel} adapted this paradigm to FB-PnP, alternating classical PnP iterations with coarse-to-fine multilevel iterations, relying on a first order coherence that can be challenging to implement.

To simplify its implementation, it has been shown in~\cite{briceno-arias_flexible_2025} that in the context of nonsmooth convex optimization, multilevel schemes are closely related to block-coordinate (BC) methods: 
when the block structure is given by a wavelet decomposition, a multilevel Forward--Backward (FB) scheme is equivalent to a block-coordinate FB scheme with a specific coarse-to-fine block activation schedule. 

More generally, BC methods offer a flexible way to mitigate the cost of large-scale problems by updating only a subset of variables at each iteration~\cite{nesterov_efficiency_2012, wright_coordinate_2015}, and several works propose block-coordinate PnP variants with blocks defined in the pixel domain. Sun et al.~\cite{sun_block_2019} introduce BC RED with i.i.d. or cyclic random block sampling; Huang et al.~\cite{huang2025deep} propose an inertial block proximal linearised scheme (iBPLM) unifying Jacobi and Gauss-Seidel-type updates for dictionary-learning blocks; Porta et al.~\cite{porta_block-coordinate_2026} reduce the memory overhead of Gradient Step denoisers~\cite{hurault_gradient_2022} with Block-PHILA, evaluating a cropped denoiser on patches \textit{via} a cyclic schedule.

If block updates in the image domain enable handling large data, 
wavelet-block activation significantly accelerates the process. Thus, we define blocks directly in the wavelet domain, denoised by a lightweight, level-specific network, and activate them adaptively \textit{via} a stochastic Gauss-Southwell-inspired rule. In the present contribution we build upon our previous work \texttt{MAGIC-FB}~\cite{desainte2026multiresolution}, which operates in a non-smooth variational context with $\ell_1$ wavelet prior. The extension of these ideas to a PnP scheme ensures better reconstructions, while keeping the effectiveness across degradation regimes granted by the adaptive selection rule.
Our contributions are threefold: 
\begin{enumerate}[label=(\roman*), itemsep=1pt, topsep=2pt]
    \item \texttt{MAGIC-FB-PnP}: a multiresolution BC PnP algorithm with wavelet-domain blocks, where PnP updates are applied to an adaptively selected subset,
    \item the design of lightweight, scale-specific denoiser networks trained jointly on approximation and detail coefficients for blockwise application,
    \item numerical tests on image deblurring showing that, among wavelet block-coordinate PnP methods, the adaptive rule matches or outperforms the best fixed block-selection strategy on different degradation regimes.
\end{enumerate}

\section{Proposed method}
\label{sec:method}

We consider a forward model of the form $ \mathbf{y} =  \mathbf{A} \bar{\mathbf{x}} +  \bm{\eta}$, where $\bar{ \mathbf{x}} \in \mathbb{R}^n$ is the unknown image, $ \mathbf{A}$ is a known linear degradation operator, and $\bm{\eta}$ is an additive Gaussian noise, centered with variance $\sigma^2$.
We aim at recovering an estimation of $\bar{\mathbf{x}}$.

The classical FB-PnP iteration reads
$$ \mathbf{x}^{[k+1]} = \mathbf{D}_\theta( \mathbf{x}^{[k]} - \tau \nabla f( \mathbf{x}^{[k]})),$$
where the gradient descent step with $f = \frac{1}{2}\|\mathbf{A}\cdot- \mathbf{y}\|^2$, enforces consistency with the forward model, while $\mathbf{D}_\theta$ is a trained denoiser that enforces an implicit prior on the solution, and $\tau > 0$ is a step size parameter.

\subsection{Block-Coordinate FB-PnP framework}
\label{sec:bc-pnp}

Given a $J$-level orthonormal wavelet transform operator $\mathbf{W} \in \mathbb R^{n \times n}$, an image $\mathbf{x}$ can be decomposed into $J+1$ blocks $ \mathbf{W}\mathbf{x} = (\mathbf{a}_J^\top, \mathbf{d}_J^\top, \ldots, \mathbf{d}_1^\top)^\top$, where $\mathbf{a}_J$ is the approximation coefficient, and $(\mathbf{d}_j)_{1 \leq j \leq J}$ are the detail coefficients, where $j=1$ corresponds to the finest scale, and $j=J$ corresponds to the coarsest scale. The three orientations (horizontal, vertical, diagonal) are stacked onto the channel dimension for each detail level, so that $\mathbf{d}_j \in \mathbb R^{3\times \tfrac{n}{4^j}}$. 
We denote $(\mathbf{w}_0^\top, \ldots, \mathbf{w}_J^\top)^\top := (\mathbf{a}_J^\top, \mathbf{d}_J^\top, \ldots, \mathbf{d}_1^\top)^\top = \mathbf{W}\mathbf{x}$ to emphasize this block structure without distinguishing approximation and detail coefficients.

We propose a BC wavelet-PnP algorithm exploiting such a decomposition where each block is denoised using a pre-trained network $\mathbf{D}_{\hat{\theta}_{\mathbf{w}_i}}$.
The iterations read, for $\tau>0$:
\begin{align}
\label{alg:wav-PnP}
    &\text{For } k=0,1,\dots \nonumber\\
    &\left|
    \begin{aligned}
    \;\; \mathbf{w}_i^{[k+1]} &= \mathbf{D}_{\hat{\theta}_{\mathbf{w}_i}}\big(\mathbf{w}_i^{[k]} - \tau \nabla_{\mathbf{w}_i} f(\mathbf{x}^{[k]})\big) & \forall i \in \{0,\ldots, J\}\\
    \mathbf{x}^{[k+1]} &= \mathbf{W}^{-1} \mathbf{w}^{[k+1]}.&
    \end{aligned}
    \right.\raisetag{17pt}
\end{align}

When the denoisers are chosen as the proximal operator of a regularization function $\tau R$~\cite{chouzenoux_block_2016, briceno-arias_flexible_2025}, the sequence $(\mathbf{x}^{[k]})_{k\in \mathbb{N}}$ converges to the solution 
$\widehat{\mathbf{x}} \in \Argmin_{\mathbf{x}\in \mathbb R^n} \frac{1}{2}\|\mathbf{A}\mathbf{x} - \mathbf{y}\|^2 + R(\mathbf{W}\mathbf{x}). $
If $R=\lambda \|\cdot\|_1$, the iterates converge to a fixed point of the classical $\ell_1$-wavelet regularized problem.

\subsection{Adaptive selection of the updated blocks}
\label{sec:adaptive-selection}

Motivated by the benefits of adaptive block-selection in the case of $\ell_1$-regularized problems \cite{desainte2026multiresolution}, we extend this selection rule to the PnP case.

Given the current iterate $\mathbf{x}^{[k]}$, and denoting by $\mathbf{s}_i^{[k]} = \mathbf{w}_i^{[k]} - \tau \nabla_{\mathbf{w}_i} f(\mathbf{x}^{[k]})$ the forward step on block $i$, we define the $i$-th denoising residual
\begin{equation}
\label{eq:pnp-residual}
\bm{\Delta}_i^{[k]} = \mathbf{w}_i^{[k]} - \mathbf{D}_{\hat{\theta}_{\mathbf{w}_i}}(\mathbf{s}_i^{[k]}),
\end{equation}
which measures how much each block would improve the restoration if it was updated.

Activation probabilities are then defined as $p_i^{[k]} = \tfrac{\|\bm{\Delta}_i^{[k]}\|}{\|\bm{\Delta}^{[k]}\|}$, where $\|\bm{\Delta}^{[k]}\|^2 = \sum_i \|\bm{\Delta}_i^{[k]}\|^2$, so that the blocks carrying the largest update are the most likely to be selected.
Finally, $J+1$ Bernoulli variables $\varepsilon_i^{[k]} \sim \mathcal B(p_i^{[k]})$ are sampled independently, and the associated activation set is $S^{[k]} = \{i : \varepsilon_i^{[k]} = 1\}$: only the activated blocks are updated, the others are left unchanged.
Algorithm~\ref{alg:bc-pnp} summarizes the resulting scheme.

Choosing every $ \mathbf{D}_{\hat{\theta}_{\mathbf{w}_i}}$ as the proximity operator of an $\ell_1$ norm, \textit{i.e.}, a soft-thresholding, 
we recover exactly the selection rule of \texttt{MAGIC-FB} \cite{desainte2026multiresolution}.
With this extension to learned denoisers we introduce a new method, which we call \texttt{MAGIC-FB-PnP}.

Selecting blocks through \eqref{eq:pnp-residual} requires computing the residual of every block, which may seem to cancel the benefit of restricting the update to a subset of them.
For the gradient step, the $\mathbf{s}_i^{[k]}$ can be cheaply computed since only the gradient terms of blocks in $S^{[k]}$ need to be recomputed from one iteration to the following one.
The backward step differs from~\cite{desainte2026multiresolution}: while in \texttt{MAGIC-FB} the denoisers are cheap soft-thresholding operations, here they are networks, so forming the residuals amounts to a full-resolution denoising pass, regardless of $S^{[k]}$.
This pass is however not wasted and used in the update, since $\mathbf{w}_i^{[k+1]} = \mathbf{w}_i^{[k]} - \bm{\Delta}_i^{[k]}$ for the activated blocks.

An iteration thus costs a fixed denoising pass plus a forward step scaling with $|S^{[k]}|$: with lightweight networks, the fixed cost stays small enough for adaptivity to reduce both per-iteration cost and iteration count, whereas larger denoisers make the fixed backward pass dominate, shrinking the wall-clock benefit to the reduction in iteration count alone.

\subsection{Wavelet block-coordinate denoiser}
\label{sec:wavelet_denoiser}

For the proposed \texttt{MAGIC-FB-PnP} method, an off-the-shelf denoiser cannot directly operate on the wavelet coefficients, as detail coefficients have a different structure to that of natural images.
We instead train a dedicated lightweight network for each wavelet block: an approximation network $\mathbf{D}_{\hat{\theta}_{\mathbf{w}_0}}$ for the coarse band, and one detail network $\mathbf{D}_{\hat{\theta}_{\mathbf{w}_i}}$ per decomposition level $i \in \{1, \ldots, J\}$.
$\mathbf{D}_{\hat{\theta}_{\mathbf{w}_i}}$ is applied independently to each of the 
 orientations, with weights shared across orientations but not across levels.

\begin{figure}
    \centering
    \includegraphics[width=0.48\textwidth]{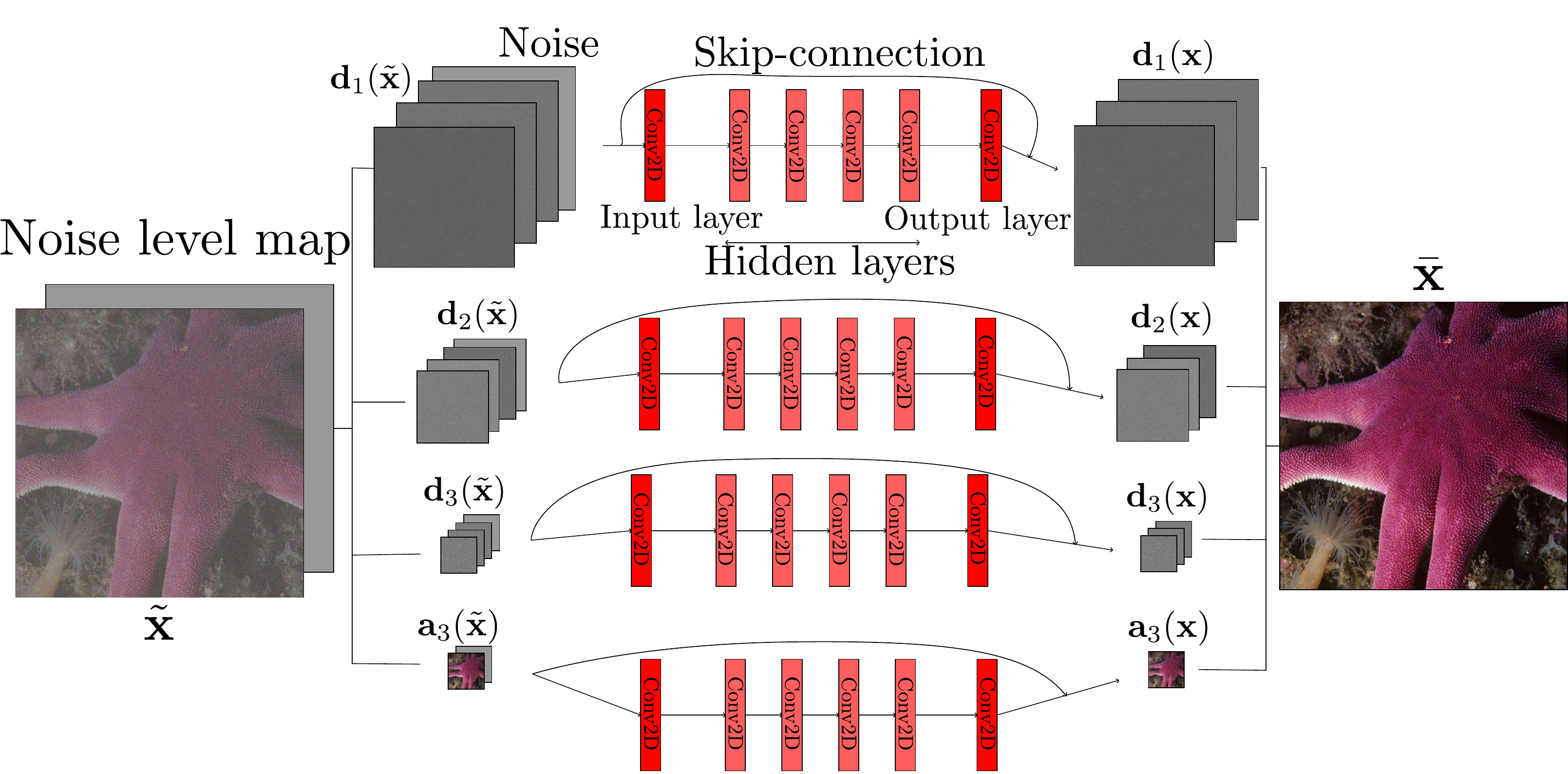}
    \caption{Block-separable wavelet denoiser. Each level shares the same architecture: a stack of convolutional layers with a skip-connection.}
    \label{fig:wavelet_denoiser}
\end{figure}

All networks $\left(\mathbf{D}_{\hat{\theta}_{\mathbf{w}_i}}\right)_{0 \leq i \leq J}$ share the same architecture, detailed on Figure~\ref{fig:wavelet_denoiser}: a stack of convolutional layers with ReLU activations, taking as input the noisy block concatenated with a constant channel filled with $\sigma$, encoding the noise level.
Each network uses a skip connection with weight $\tfrac{1}{2}$: it returns the average of its input and the raw output of the convolutional stack, rather than the raw output directly.

We train all networks jointly by minimizing an $\ell_2$ loss on all the approximation and detail coefficients.
Given $\overline{\mathbf{x}}$ a clean training image, $\widetilde{\mathbf{x}} = \overline{\mathbf{x}} + \bm{\varepsilon}$ a noisy version of it with $\bm{\varepsilon} \sim \mathcal{N}(0, \tilde{\sigma}^2 I)$ and $\tilde{\sigma} \sim p_{\tilde{\sigma}}$ log-uniform on $[\sigma_{\min}, \sigma_{\max}]$, we define  
\begin{equation}
    \label{eq:loss} 
    \mathcal{L}(\theta_{\mathbf{w}_0}, \ldots, \theta_{\mathbf{w}_J}) = \mathbb{E}\bigg[\bigg\| \mathbf{W}\overline{\mathbf{x}} - \begin{pmatrix} \mathbf{D}_{\theta_{\mathbf{w}_0}} \\ \vdots \\ \mathbf{D}_{\theta_{\mathbf{w}_J}}
    \end{pmatrix}(\mathbf{W} \Tilde{\mathbf{x}})
    \bigg\|^2\bigg],
\end{equation}
where the expectation is taken over $\overline{\mathbf{x}}$, $\bm{\varepsilon}$, and $\tilde{\sigma}$ and the vector of networks acts block-wise, each network being applied to the block of $\mathbf{W}\widetilde{\mathbf{x}}$ it is associated with.
The noise is therefore sampled in the image domain and only then decomposed.
We denote with $\hat{\bm{\theta}} = (\hat\theta_{\mathbf{w}_0}, \ldots, \hat\theta_{\mathbf{w}_J}) \in \operatorname*{Argmin} \mathcal{L}$ the resulting trained parameters, used throughout the paper.
Orthonormality makes \eqref{eq:loss} equal to the image-domain loss $\mathbb{E}\|\overline{\mathbf{x}} - \mathbf{W}^{-1}(\mathbf{D}_{\theta_{\mathbf{w}_0}}, \ldots, \mathbf{D}_{\theta_{\mathbf{w}_J}})(\mathbf{W}\tilde{\mathbf{x}})\|^2$, so the block-separable denoiser is trained exactly as an image denoiser.

\begin{algorithm}[t]
\caption{\texttt{MAGIC-FB-PnP}}
\label{alg:bc-pnp}
\begin{algorithmic}
\State \textbf{Input:} initial point $\mathbf{w}^{[0]} = (\mathbf{w}_0^{[0]\top}, \ldots, \mathbf{w}_J^{[0]\top})^\top$, noise level $\sigma$, step size $\tau$
\For{$k = 0, 1, \ldots$}
\For{$i = 0, \ldots, J$}
\State $\mathbf{s}_i^{[k]} = \mathbf{w}_i^{[k]} - \tau \nabla_{\mathbf{w}_i} f(\mathbf{x}^{[k]})$
\State $\bm{\Delta}_i^{[k]} = \mathbf{w}_i^{[k]} - \mathbf{D}_{\hat{\theta}_{\mathbf{w}_i}}(\mathbf{s}_i^{[k]})$
\EndFor
\State Draw $\varepsilon_i^{[k]} \sim \mathcal B(p_i^{[k]})$ with $p_i^{[k]} = \|\bm{\Delta}_i^{[k]}\| / \|\bm{\Delta}^{[k]}\|$, and set $S^{[k]} = \{i : \varepsilon_i^{[k]} = 1\}$
\For{$i = 0, \ldots, J$}
\If{$i \in S^{[k]}$}
\State $\mathbf{w}_i^{[k+1]} = \mathbf{D}_{\hat{\theta}_{\mathbf{w}_i}}(\mathbf{s}_i^{[k]})$
\Else
\State $\mathbf{w}_i^{[k+1]} = \mathbf{w}_i^{[k]}$
\EndIf
\EndFor
\State $\mathbf{x}^{[k+1]} = \mathbf{W}^{-1} \mathbf{w}^{[k+1]}$
\EndFor
\end{algorithmic}
\end{algorithm}

\subsection{Convergence guarantees}
\label{sec:convergence}

\begin{proposition}
\label{prop:cv}
Let $\tau \in (0, 2/\|\mathbf{A}\|^2)$ and assume that every block denoiser $\mathbf{D}_{\hat{\theta}_{\mathbf{w}_i}}$ is $\tfrac{1}{2}$-averaged, so that the operator $\mathbf{T}$ defined blockwise by $\mathbf{T}_i(\mathbf{x}) = \mathbf{D}_{\hat{\theta}_{\mathbf{w}_i}}(\mathbf{w}_i - \tau \nabla_{\mathbf{w}_i} f(\mathbf{x}))$ is averaged\footnote{See for example \cite{bauschke_convex_2017}.} and satisfies $\mathrm{fix}(\mathbf{T}) \neq \emptyset$.
Assume  that the activation probabilities of Section~\ref{sec:adaptive-selection} satisfy $p_i^{[k]} / p_i^{[k+1]} \leq 1 + \eta_k$ for a summable sequence $(\eta_k)_k$, and $p_i^{[k]} \geq p_{\min} > 0$ for every $i$ and $k$.
Then $\bm{\Delta}^{[k]} \to 0$ and the sequence $(\mathbf{x}^{[k]})_k$ generated by Algorithm~\ref{alg:bc-pnp} converges almost surely to a fixed point of $\mathbf{T}$.
\end{proposition}

The proof relies on the variable-metric quasi-Féjér framework of \cite{combettes_variable_2013}: the averagedness of $\mathbf{T}$ yields a contraction in the metric induced by the activation probabilities $(p_i^{[k]})_i$, and $p_i^{[k]}/p_i^{[k+1]} \leq 1+\eta_k$ ensures this metric varies slowly enough for the Robbins--Siegmund lemma~\cite{robbins1971convergence} to apply. A full proof will be given in a forthcoming extended version.

\section{Numerical results}
\label{sec:results}

\begin{figure*}[t]
    \centering
    \begin{subfigure}[b]{0.49\textwidth}
        \centering
        \caption{Low blur, high noise ($\sigma_{\text{blur}} = 3.0$, $\sigma = 0.1$)}
        \includegraphics[width=\textwidth]{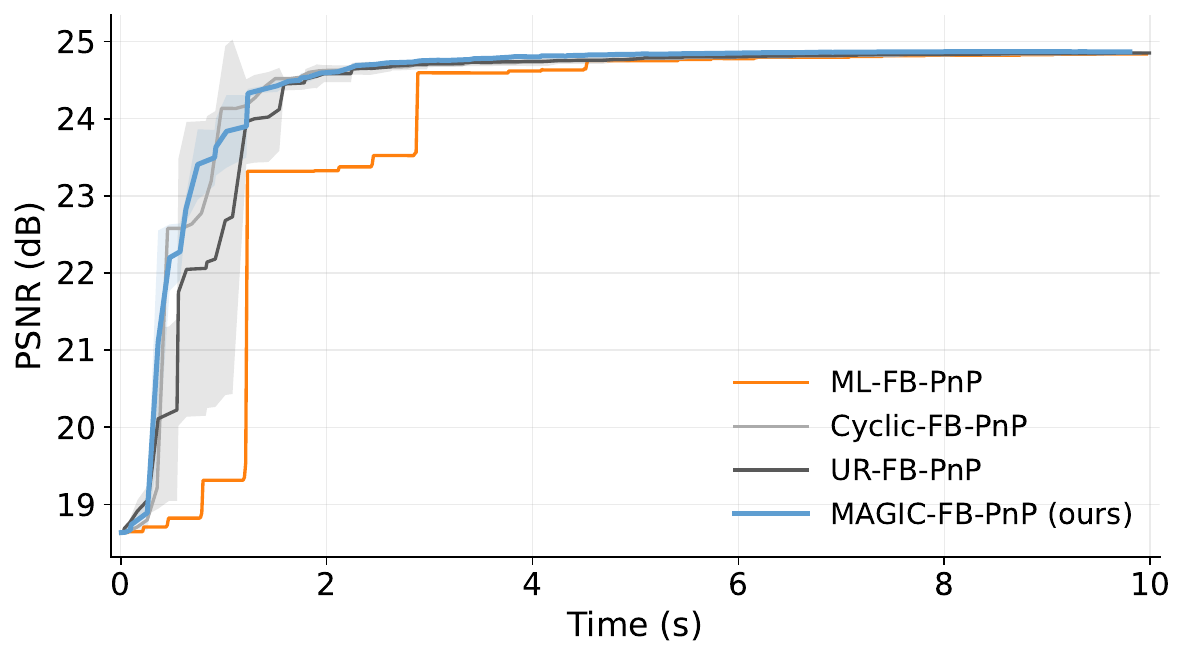}
        \label{fig:psnr-time:lowblur}
    \end{subfigure}
    \hfill
    \begin{subfigure}[b]{0.49\textwidth}
        \centering
        \caption{High blur, low noise ($\sigma_{\text{blur}} = 10.0$, $\sigma = 0.01$)}
        \includegraphics[width=\textwidth]{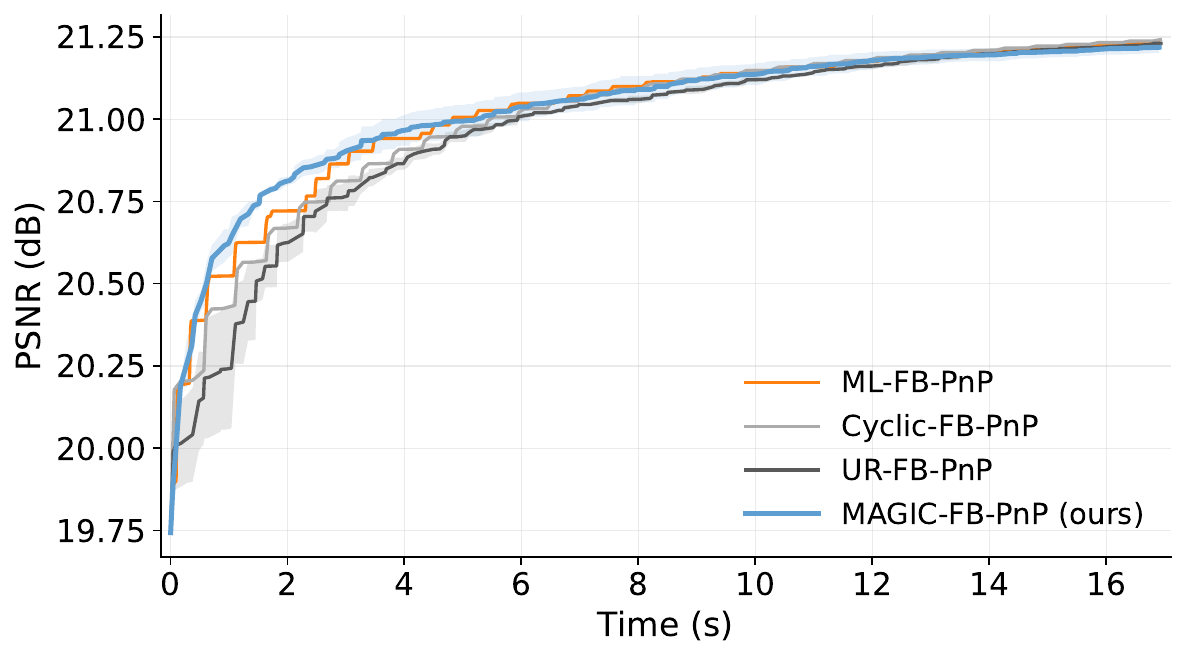}
        \label{fig:psnr-time:highblur}
    \end{subfigure}
    \vspace{-0.5cm}
    \caption{PSNR against wall-clock time. Shaded areas report $\pm 1$ standard deviation over 5 seeds.}
    \label{fig:psnr-time}
\end{figure*}

\begin{figure*}[t]
    \centering
    \begin{subfigure}[b]{0.24\textwidth}
        \centering
        \includegraphics[width=0.95\linewidth]{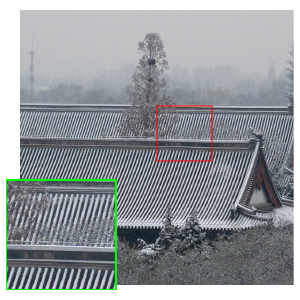}
        \caption*{Ground truth\\[-1pt] \footnotesize \hspace{0.1cm}\\[-1pt] \footnotesize \hspace{0.1cm}}
        \label{fig:recon:xtrue}
    \end{subfigure}%
    \hspace{0.005\textwidth}%
    \begin{subfigure}[b]{0.24\textwidth}
        \centering
        \includegraphics[width=0.95\linewidth]{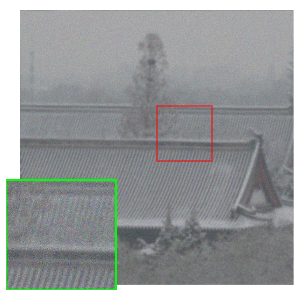}
        \caption*{Observation\\[-1pt] \footnotesize PSNR: 17.90 dB \\ SSIM: 0.145}
        \label{fig:recon:y}
    \end{subfigure}%
    \hspace{0.005\textwidth}%
    \begin{subfigure}[b]{0.24\textwidth}
        \centering
        \includegraphics[width=0.95\linewidth]{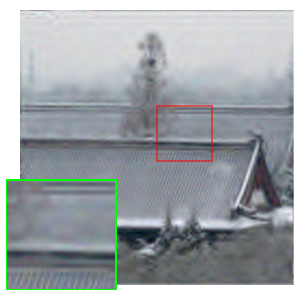}
        \caption*{$\ell_1$ wavelet\\[-1pt] \footnotesize PSNR: 20.35 dB \\ SSIM: 0.466}
        \label{fig:recon:l1}
    \end{subfigure}%
    \hspace{0.005\textwidth}%
    \begin{subfigure}[b]{0.24\textwidth}
        \centering
        \includegraphics[width=0.95\linewidth]{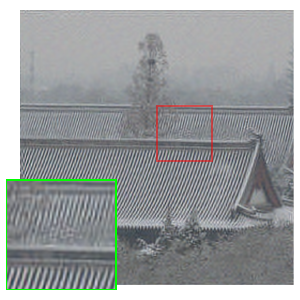}
        \caption*{\texttt{MAGIC-FB-PnP}\\[-1pt] \footnotesize PSNR: 24.97 dB \\ SSIM: 0.679}
        \label{fig:recon:magic}
    \end{subfigure}
    \caption{Reconstruction in the low blur, high noise setting ($\sigma_{\text{blur}} = 3.0$, $\sigma = 0.1$).}
    \label{fig:recon}
\end{figure*}

We consider Gaussian deblurring, $\mathbf{y} = \mathbf{A}\bar{\mathbf{x}} + \bm{\eta}$, with $\mathbf{A}$ a blur of standard deviation $\sigma_{\text{blur}}$ and $\bm{\eta} \sim \mathcal{N}(0, \sigma^2 \mathbf{I})$, on $1024\times 1024$ images from the LIU4K dataset~\cite{LIU4K}.
We consider two degradation regimes: low blur, high noise ($\sigma_{\text{blur}} = 3.0$, $\sigma = 0.1$) and high blur, low noise ($\sigma_{\text{blur}} = 10.0$, $\sigma = 0.01$).
We use a $J=5$-level Daubechies (db8) decomposition, and results for stochastic strategies are averaged over 5 seeds.

\subsection{Denoiser training}
Each block network $\mathbf{D}_{\hat{\theta}_{\mathbf{w}_i}}$ is a 6-layer, 64-channel CNN with $3\times 3$ convolutions and ReLU activations, taking as input the noisy coefficients concatenated with a constant channel encoding $\tilde\sigma$.
The $J+1=6$ networks (around 1M parameters in total, versus 32M for DRUNet~\cite{zhang2021plug}) are trained jointly on DIV2K by minimizing \eqref{eq:loss} with Adam (learning rate $10^{-4}$, weight decay $10^{-4}$, gradient clipping at 1, batch size 2, cosine-annealing schedule, 1000 epochs), with $\tilde\sigma$ sampled log-uniformly in $[0.01, 0.5]$ for every training pair.

\subsection{Block activation strategies}
We compare the proposed adaptive rule (\texttt{MAGIC-FB-PnP}) against three fixed block-selection strategies, all instances of~\eqref{alg:wav-PnP} using the same wavelet block-coordinate denoiser:
\begin{itemize}
    \item \texttt{Cyclic-FB-PnP}: a fixed cyclic schedule, activating one block at a time in rotation.
    \item \texttt{ML-FB-PnP}: a deterministic coarse-to-fine multilevel schedule (Multilevel Plug-and-Play~\cite{laurent2025multilevel}), activating a growing set of blocks $\{\mathbf{a}_J\}$, $\{\mathbf{a}_J, \mathbf{d}_J\}$, $\{\mathbf{a}_J, \mathbf{d}_J, \mathbf{d}_{J-1}\}, \ldots$ until all blocks are active, then restarting from $\{\mathbf{a}_J\}$.
    \item \texttt{UR-FB-PnP}: i.i.d. uniform random block sampling, where each block has a $50\%$ chance of being updated.
\end{itemize}

The behaviour of the classical FB-PnP that updates all blocks at each iteration is close to that of \texttt{Cyclic-FB-PnP}, and is thus not reported in the figure.

\subsection{Results and discussion}

Figure~\ref{fig:psnr-time} reports the PSNR trajectories in both regimes.
In the high blur, low noise setting, \texttt{ML-FB-PnP} and \texttt{MAGIC-FB-PnP} converge faster than \texttt{Cyclic-FB-PnP} and \texttt{UR-FB-PnP}.
This is consistent with the multilevel literature, where coarse updates carry most of the progress when the operator $A$ strongly attenuates high frequencies.
In the low blur, high noise setting, the ordering changes: the fixed coarse-to-fine schedule no longer focuses on the most impactful blocks, and \texttt{MAGIC-FB-PnP} reaches convergence faster than \texttt{ML-FB-PnP}.
\texttt{Cyclic-FB-PnP} and \texttt{UR-FB-PnP} work best in this setting, where high-frequencies need to be updated faster to improve convergence speed.

\texttt{MAGIC-FB-PnP} is thus the fastest, or matches the fastest method, in both regimes, whereas each fixed schedule is competitive in only one of them.
This is the practical benefit of the Gauss--Southwell inspired selection: the activation pattern is recovered from the iterates themselves, without having to identify the degradation regime beforehand and select a schedule accordingly.

Finally, Figure~\ref{fig:recon} illustrates the gain brought by the learned block denoisers over the $\ell_1$ wavelet sparsity prior used in \cite{desainte2026multiresolution} on the same block-coordinate scheme, with an improvement of 4.62 dB in PSNR and 0.213 in SSIM.

\section{Conclusion}
\label{sec:conclusion}
We proposed a multiresolution block-coordinate FB-PnP algorithm, extending the adaptive Gauss--Southwell inspired selection rule in \cite{desainte2026multiresolution} to learned block denoisers, with convergence guaranteed under an averagedness assumption, and experiments on Gaussian deblurring showing that it matches the fastest fixed activation schedule across degradation regimes without prior knowledge on the regime.

As discussed in Section~\ref{sec:adaptive-selection}, computing the activation probabilities still requires a full-resolution denoising pass at every iteration. A cheaper proxy for block selection is a natural next step.
More broadly, better exploiting the multiresolution structure, \textit{e.g.} \textit{via} scaling laws linking denoiser capacity to wavelet scale, could be a promising direction.

\section*{Acknowledgment}
This work was funded by the ANR-24-CE23-7039 MEPHISTO and ANR-26-CE48-5044-01 DeepScale projects, the Simone and Cino Del Duca foundation and the PEPR IA project. We thank the Blaise Pascal Center for its computational support, using the SIDUS \cite{quemener2013sidus} solution.

\printbibliography

\end{document}